\documentclass[10pt,aps,twocolumn,prd,superscriptaddress,showpacs,nofootinbib,noshowkeys,floatfix,preprintnumbers]{revtex4}
\usepackage{graphics,graphicx}
\usepackage{amsmath, amssymb}
\usepackage{multirow}
\usepackage{longtable}
\usepackage{color}

\usepackage[normalem]{ulem}  

\begin{document}
\title{Polyakov loop fluctuations in SU(3) lattice gauge theory
\\
and an effective gluon potential
}
\date{\today}
\author{Pok Man Lo}
\affiliation{GSI, Helmholzzentrum f\"{u}r Schwerionenforschung,
Planckstrasse 1, D-64291 Darmstadt, Germany}
\author{Bengt Friman}
\affiliation{GSI, Helmholzzentrum f\"{u}r Schwerionenforschung,
Planckstrasse 1, D-64291 Darmstadt, Germany}
\author{Olaf Kaczmarek}
\affiliation{Fakult\"at f\"ur Physik, Universit\"at Bielefeld, 33615 Bielefeld, Germany}
\author{Krzysztof Redlich}
\affiliation{Institute of Theoretical Physics, University of Wroclaw,
PL-50204 Wroc\l aw, Poland}
\affiliation{Extreme Matter Institute EMMI, GSI,
Planckstrasse 1, D-64291 Darmstadt, Germany}
\author{Chihiro Sasaki}
\affiliation{Frankfurt Institute for Advanced Studies, D-60438 Frankfurt am Main, Germany}

\begin{abstract}
 {
We calculate the Polyakov loop susceptibilities in the SU(3) lattice gauge theory using
the  Symanzik improved gauge action on different-sized lattices.
The longitudinal and transverse fluctuations of the Polyakov loop, as well as, that of its absolute value are considered. We analyze their 
  properties
 in relation  to the confinement-deconfinement phase transition.

 We also present results based on simulations of (2+1)-flavor QCD on
  $32^3\times 8$
  lattice using Highly Improved Staggered Quark (HISQ) action by the HotQCD collaboration. The influences of fermions on the Polyakov loop fluctuations are discussed. 
We show, that ratios of different susceptibilities of the Polyakov loop are sensitive  probes for critical behavior.
   We formulate an effective model for the
Polyakov loop  potential
 and constrain  its parameters from  existing quenched lattice data including fluctuations. We emphasize the role of fluctuations to fully explore the thermodynamics of pure gauge theory within an effective approach.
}
\end{abstract}
\maketitle

\section{Introduction}

Lattice gauge theory provides a powerful nonperturbative approach to study the phase structure and thermodynamics of a non-abelian  gauge theory.

Presumably,   the best explored till now
are pure lattice gauge theories with the SU(2) and SU(3) color groups \cite{su21,su22,Engels:1994xj,Engels:1998nv,Boyd:1995zg,su31}.
Recently, the equation of state of the SU(3) gauge theory was established with an
unprecedented precision in a broad parameter range within lattice approach \cite{Borsanyi:2012ve}. The results were extrapolated to the continuum and thermodynamic limit, providing definitive results on the equation of state and the value of critical temperature.

The phase transition in the SU(3) gauge theory is first order and is characterized by the spontaneous breaking of the global $\mathcal{Z}_3$ center symmetry of the Yang-Mills Lagrangian.  The Polyakov loop is an order parameter which is linked to the free energy of the static quark immersed in a hot gluonic medium \cite{McLerran:1980pk,Kaczmarek:2002mc}.
 At low temperatures the   thermal expectation value of the Polyakov loop   $\langle | L|\rangle $ vanishes,  indicating color confinement. At high temperatures  $\langle | L|\rangle\neq  0$, resulting in the finite energy of a static quark, and consequently deconfinement of color and the spontaneous breaking of $\mathcal{Z}_3$ center symmetry.
The transition temperature can be identified from the peak position of the Polyakov loop susceptibility.

On the lattice, the ultraviolet divergence of the bare quark-antiquark free energy implies, that in the continuum limit, the bare Polyakov loop vanishes for all temperatures. Thus, the Polyakov loop and the related susceptibilities require renormalization to become physically meaningful \cite{Kaczmarek:2002mc, gavai}. While the basic thermodynamics functions of the SU(3) pure gauge theory are well established within lattice approach, the situation is less clear for the temperature dependence of the renormalized Polyakov loop and its susceptibilities.

In this paper we  focus on the renormalized Polyakov loop susceptibilities. For color gauge group $SU(N_c\ge 3)$, the Polyakov loop is complex-valued. Consequently, discussing the Polyakov loop fluctuations one can consider susceptibilities along longitudinal and transverse direction, as well as, that of its absolute value. \footnote{In the real sector of the Polyakov loop, longitudinal and transverse components correspond to the real and imaginary direction respectively. Although the thermal average of the imaginary part of the Polyakov loop vanishes, its fluctuation along the imaginary direction does not.}

We  calculate  the temperature dependence of the Polyakov loop
 susceptibilities within  the  SU(3) lattice gauge theory. We use
the   Symanzik improved gauge action on  $N_\sigma^3\times
N_\tau$ lattices for different   values of the temporal lattice sizes $N_\tau=(4,6,8)$   and for
spatial extensions $N_\sigma$ varying from 16 to 64.

In our recent studies \cite{prd}, we have discussed the ratios of different susceptibilities of the Polyakov loop and motivate   their importance   as  observables  to  probe deconfinement transition in the SU(3)  pure gauge theory.
 We have shown, that the ratios of susceptibilities display a  jump across the critical temperature.

In this paper we extend our previous study and discuss  properties of the Polyakov  susceptibilities in a broader  temperature range.
We also study the influence of dynamical quarks on different susceptibilities  within lattice gauge theory.  We present results based on the Polyakov loop data from simulations in (2+1)-flavor QCD on  $32^3\times 8$ lattice using the HISQ action with an almost physical strange quark mass and $m_{u,d}=m_s/20$ \cite{ejiri5, hisq}.

We show, that the  explicit breaking of the $\mathcal{Z}_3$ center symmetry in QCD, modifies the properties of  the Polyakov loop
susceptibilities found in  the pure gauge theory. The ratios are substantially smoothened but still feature interesting characteristics in connection to the deconfinement phase transition.


In the confined phase and up to near  $T_c$, the thermodynamics of the SU(3) gauge theory is well described within the Hagedorn-type  model,
incorporating glueballs as degrees of freedom \cite{Borsanyi:2012ve}. In  the deconfined  phase, the perturbative approach describes the lattice data from the Stefan-Boltzmann  limit down
to $T> (2-3)T_c$ \cite{Borsanyi:2012ve}. To quantify  the equation of state near  $T_c$ one needs in general, an intrinsic non-perturbative description in terms of effective models.

The Polyakov loop is the relevant  quantity to effectively describe  gluo-dynamics \cite{meisinger,Fukushima2004277,PhysRevD.86.014007}. Various models based on the Polyakov loop have been proposed to quantify
the deconfinement transition and thermodynamics of a pure gauge theory \cite{meisinger,Fukushima2004277,PhysRevD.86.014007,Roessner:2006xn,Sasaki:2006ww,PhysRevD.73.014019,Dumitru:2002cf,Dumitru:2003hp,Herbst:2010rf,review}

We take advantage of our new data on different susceptibilities
to construct an effective potential for the
Polyakov loop, which is consistent with all existing lattice data over a broad range of temperatures.  We show, that the incorporation of fluctuation effects, is important to   describe thermodynamics of a pure gauge theory.

The  paper is organized as follows. In the next Section, we present our lattice results for the Polyakov loop  susceptibilities and their ratios.   In Section III we introduce the  effective Polyakov loop model and  its comparison with lattice data . In the last section we present our summary and conclusions.

\section{ The Polyakov loop susceptibilities on the lattice}


On a $N_\sigma^3\times N_\tau$ lattice,  the Polyakov loop is defined as the trace of the product over temporal gauge links,

\begin{eqnarray}
        L_{\vec x}^{\rm bare} ={\frac{1}{N_c}} Tr \prod_{\tau=1}^{N_\tau} U_{(\vec x,\tau),4}\, ,\\
L^{\rm bare} = \frac{1}{N_\sigma^3}\sum_{\vec x} L_{\vec x}^{\rm bare}\, .
\end{eqnarray}

Due to the $\mathcal{Z}_3$ symmetry of the pure gauge action, this quantity is exactly zero,  when
averaging over all gauge field configurations. Furthermore, the Polyakov loop is
strongly $N_\tau$ dependent  and needs to be renormalized. We consider the multiplicatively renormalized
Polyakov loop \cite{Kaczmarek:2002mc},

\begin{eqnarray}\label{r1}
 L^{\rm ren} = \left(Z(g^2)\right)^{N_\tau} L^{\rm bare}
\end{eqnarray} and introduce the ensemble average of the modulus thereof, $ \langle \vert L^{\rm ren} \vert \rangle$. The latter is well defined in the continuum and thermodynamic limits.

The thermal average of the renormalized Polyakov loop can also be obtained from the long distance behavior of the renormalized heavy
quark-antiquark free energy $F_{q \bar q}^{\rm ren}$ \cite{Kaczmarek:2002mc},  which is defined by the correlation function
\begin{eqnarray}
\langle L^{\rm ren}_{\vec x} L^{ \dagger\rm ren}_{\vec y}\rangle &=&
e^{-F_{q \bar q}^{\rm ren}(r=\vert
  \vec x - \vec y \vert,T)/T}\nonumber \\
& \stackrel{r \to \infty}{\to} & \vert\langle L^{\rm ren} \rangle \vert^2. \label{rr1}
\end{eqnarray}

In the confined phase, the heavy quark-antiquark free energy rises linearly with distance as $r \rightarrow \infty$, hence the Polyakov loop vanishes. The Polyakov loop is finite in the deconfined phase, and can serve as an order parameter for the spontaneous breaking of the $\mathcal{Z}_3$ center symmetry. Note that both definitions, \eqref{r1} and \eqref{rr1}, are equivalent in the thermodynamic limit.

Using the renormalized Polyakov loop, one can define the renormalized Polyakov loop
susceptibility as
\begin{eqnarray}\label{eq:chiA}
T^3 \chi_A =& \frac{N_\sigma^3}{N_\tau^3} \left( \langle \vert L^{\rm ren} \vert^2 \rangle - \langle
   \vert L^{\rm ren} \vert \rangle^2\right).
\end{eqnarray}
Here, we anticipate  the  factor $T^3$ to define \eqref{eq:chiA} as an observable in the continuum limit.

In the SU(3) gauge theory, the Polyakov loop operator is complex. Thus in addition to $\chi_A$, one can also explore the longitudinal and transverse fluctuations of the Polyakov loop \cite{prd}

\begin{eqnarray}
T^3 \chi_{L} =& \frac{N_\sigma^3}{N_\tau^3} \left[ \langle  (L^{\rm ren}_{L})^2 \rangle - \langle
    L^{\rm ren}_{L} \rangle^2\right].\label{eq:chiL}\\
T^3 \chi_{T} =& \frac{N_\sigma^3}{N_\tau^3} \left[ \langle  (L^{\rm ren}_{T})^2 \rangle - \langle
    L^{\rm ren}_{T} \rangle^2\right],\label{eq:chiT}
    \end{eqnarray}

\begin{figure}[!ht]
 \includegraphics[width=3.375in]{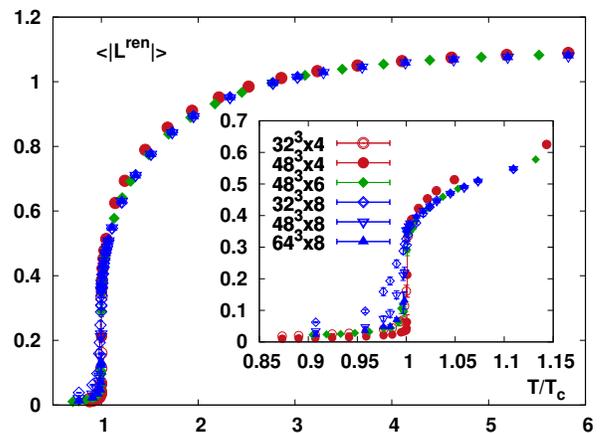}
 \caption{The temperature dependence of  the modulus of the renormalized Polyakov loop  in  the SU(3) gauge  theory, calculated on various lattices.
 } 
  \label{fig:polyakov}
\end{figure}

\begin{figure*}[!t]
 \includegraphics[width=2.31in]{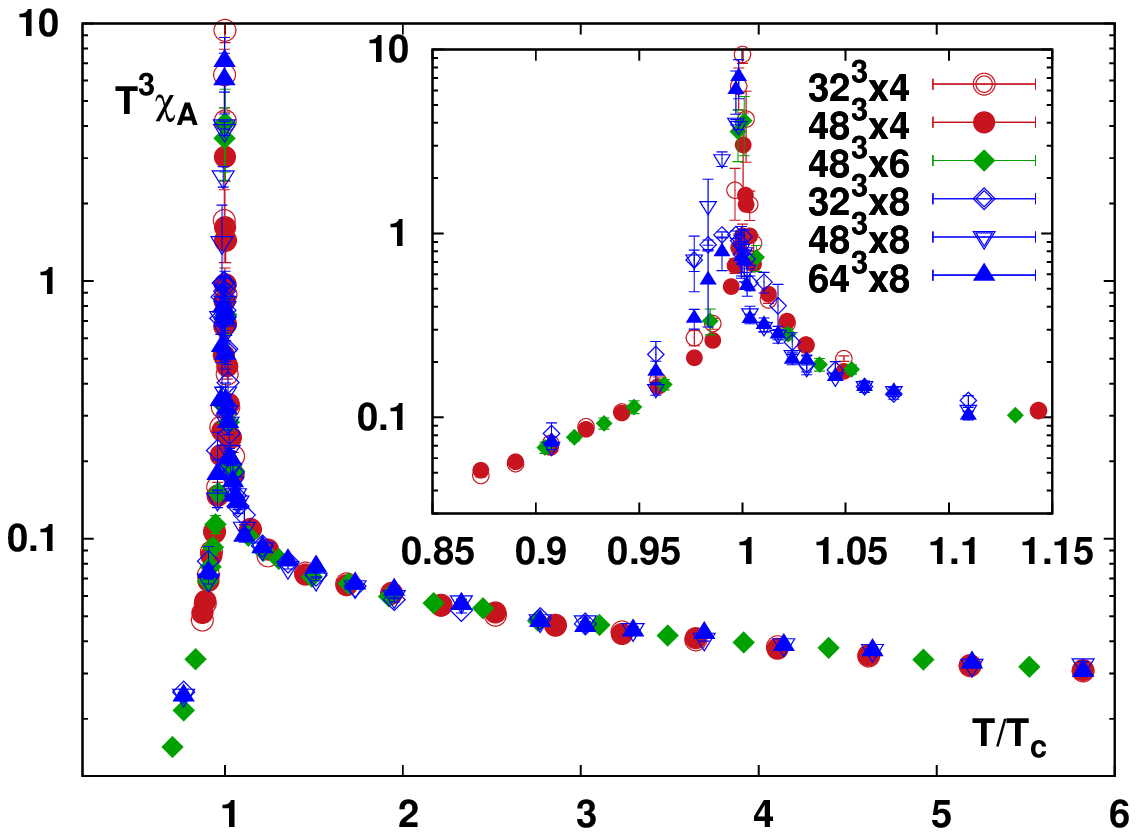}
  \includegraphics[width=2.31in]{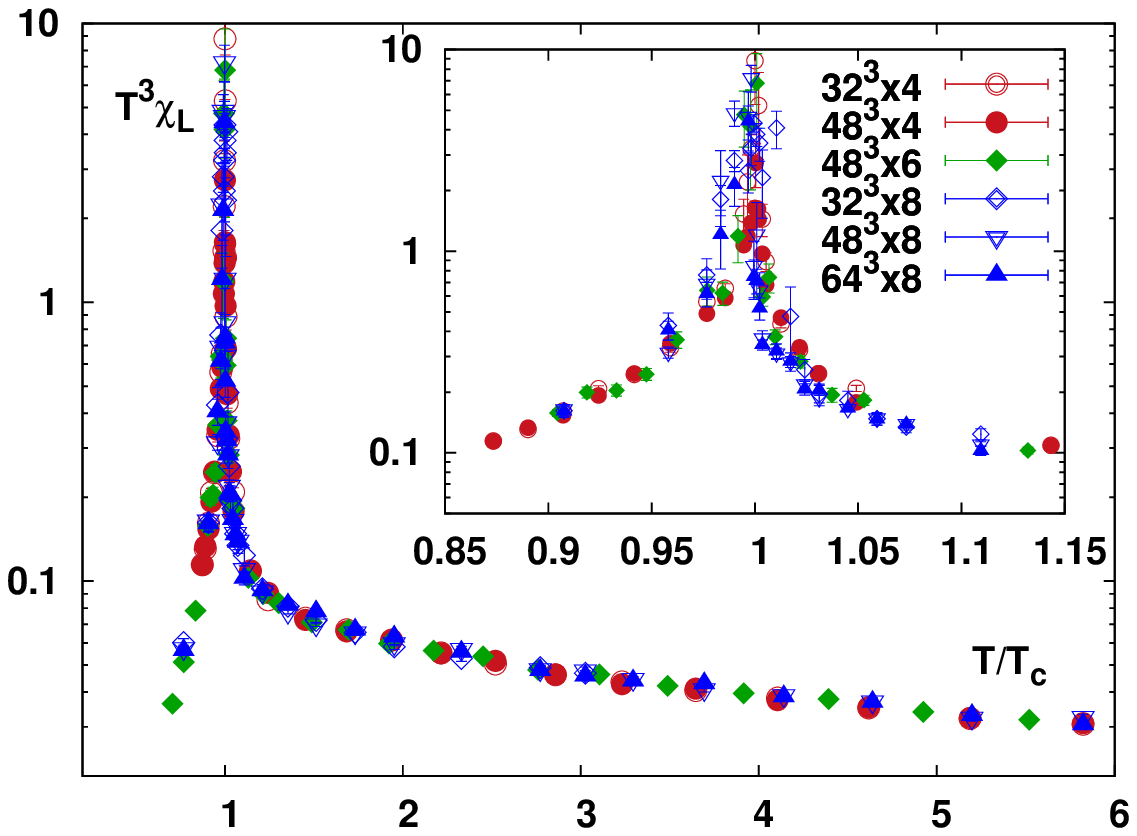}
 \includegraphics[width=2.31in]{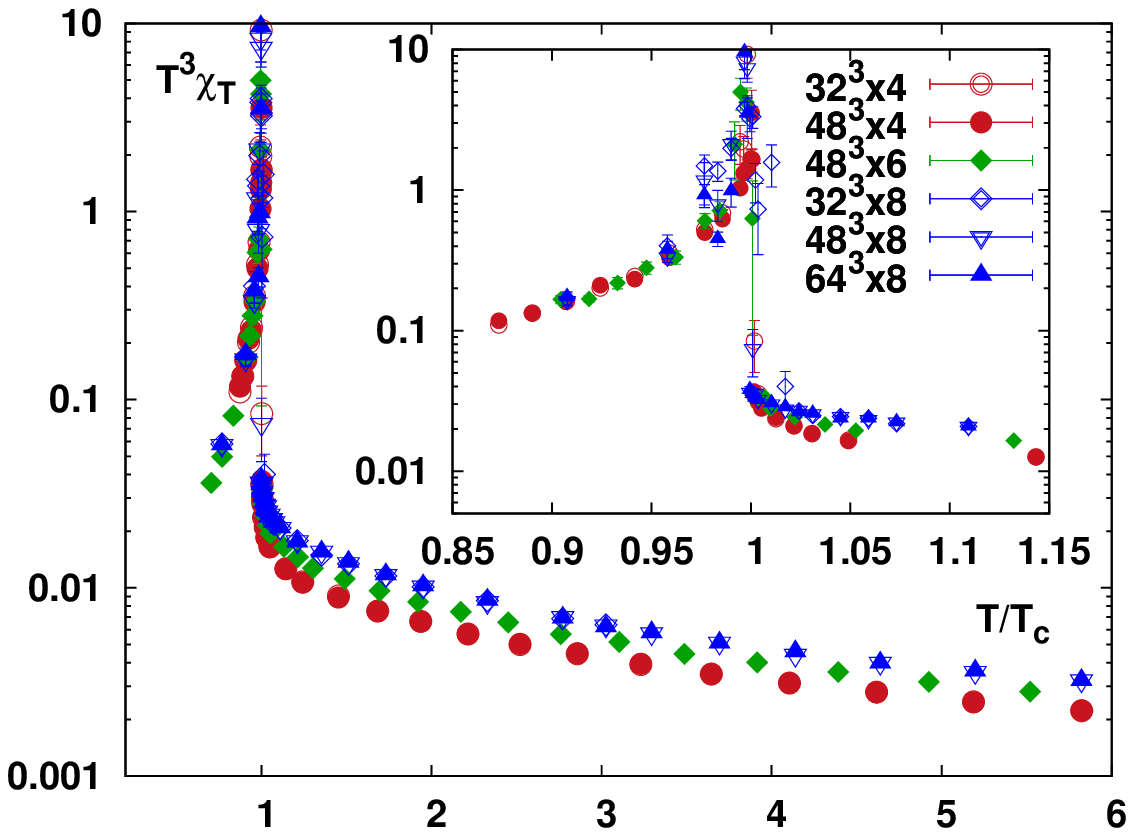}
 \caption{ The temperature dependence of the renormalized Polyakov loop susceptibilities from   Eqs. \eqref{eq:chiA}, \eqref{eq:chiL} and \eqref{eq:chiT},  calculated on various lattice sizes, in the SU(3) pure gauge theory.  The temperature is normalized to its critical value. }
 \label{fig:sus}
\end{figure*}

\begin{figure}[!ht]
 \includegraphics[width=3.355in]{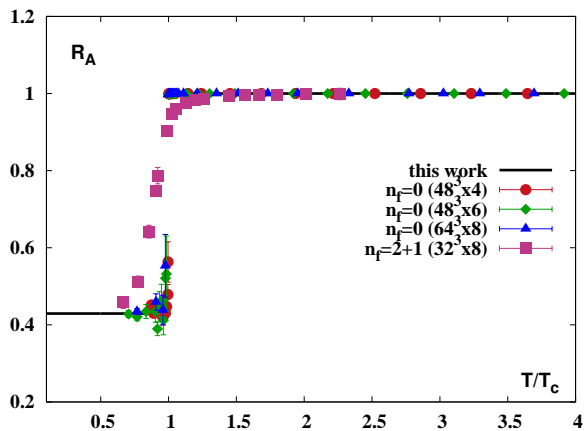}
 \caption{The ratio of the absolute \eqref{eq:chiA}  to longitudinal \eqref{eq:chiL} part of the Polyakov loop  susceptibilities  calculated  within lattice gauge theory for pure gauge system and (2+1)-flavor QCD (see text). The temperature is normalized to its (pseudo) critical value for respective lattice. The line is the model result explained in the text.  }
\label{ratioA}
\end{figure}

\begin{figure}[!ht]
\includegraphics[width=3.355in]{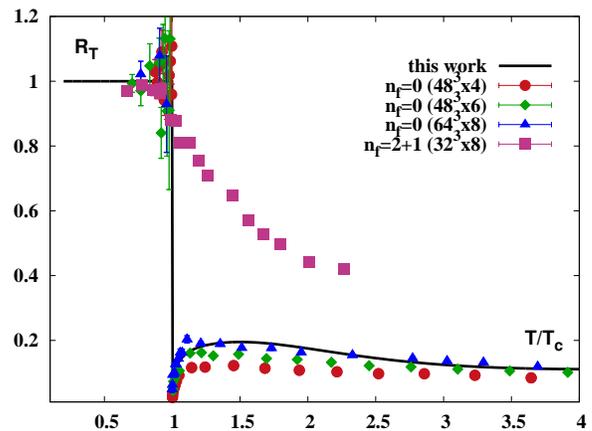}
\caption{
        Lattice results on the ratio of the transverse  \eqref{eq:chiT}  to longitudinal \eqref{eq:chiL}  susceptibility of the  Polyakov loop for pure gauge system and (2+1)-flavor QCD.
The line is the Polyakov loop model result discussed in Section III.
}
    \label{ratioT}
\end{figure}

\begin{figure*}[!t]
 \includegraphics[width=3.305in]{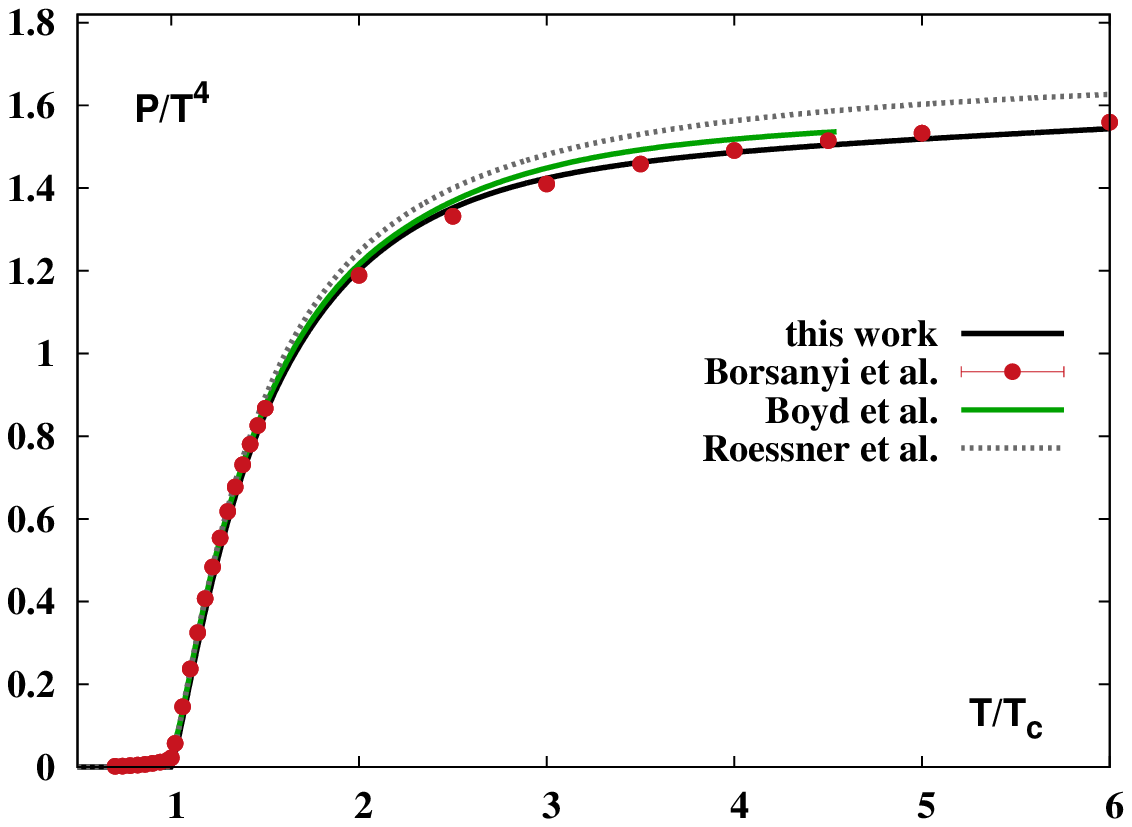}
  \includegraphics[width=3.305in] {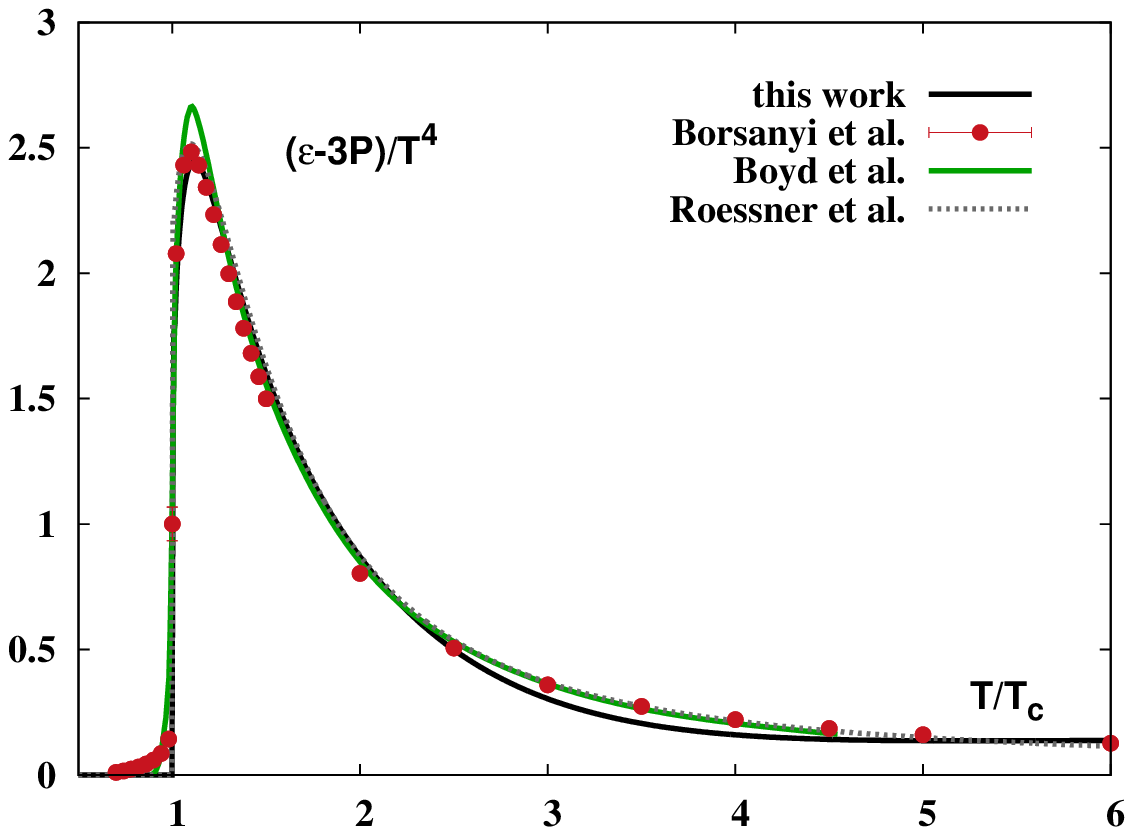}
  \caption{ Left-hand figure: Lattice QCD data for thermodynamic pressure obtained in the SU(3) gauge theory.
  The points are from Ref. \cite{Borsanyi:2012ve}, whereas the green line is the parametrization of lattice data   from Ref. \cite{Boyd:1995zg}.
   The black and the dashed lines are obtained in the Polyakov loop models introduced in Section III.
   Right-hand figure: As in  the left-hand figure but for the interaction measure $(\epsilon-3P)/T^4$, where $\epsilon$ is the energy density.
   }
 \label{fig:PM}
\end{figure*}


We have computed the Polyakov loop susceptibilities using the (1,2)-tree-level  Symanzik  improved gauge action on a $N_\sigma^3\times
N_\tau$  lattice. We consider lattices of temporal size $N_\tau=4,6$ and $8$ and spatial extent  $N_\sigma$ varying from 16 to 64.
The temperatures for the three temporal lattice extents are set by varying the bare coupling and use the
temperature scale determined by the zero temperature string tension, as well as the critical couplings of the
deconfinement transition \cite{ref3, Beinlich:1997ia}. The gauge field
configurations were generated using one heatbath and four overrelaxation
updates per sweep with $15\,000$ sweeps in general and up to $100\,000$ sweeps
close to the critical temperature, $T_c$.

The renormalization constants, $Z(g^2)$, were taken from
\cite{Kaczmarek:2002mc}. The statistical errors were obtained from a Jackknife analysis
and do not include any systematic error resulting from the renormalization procedure.
In Fig. \ref{fig:polyakov}, we show the lattice gauge theory result for $ \langle \vert L^{\rm ren} \vert \rangle$ as a function of temperature.

While no volume effects are visible in the deconfined
phase, data at fixed $N_\tau$ in the confined phase, show the expected $1/\sqrt V$
volume-dependence.
Considering
results at fixed  ratio $N_\sigma/N_\tau$, only small cut-off effects
can be observed at high as well as at low temperatures. The deviation of the
$N_\tau=4$ and 8 data between $(1-2) T_c$ may  be attributed to
the uncertainty in the determination of the renormalization constants, rather
than to the cut-off effects.

The results for the renormalized Polyakov loop susceptibilities obtained on different lattice
sizes are shown in Fig.~\ref{fig:sus}. In the close vicinity of the phase
transition, $0.95 < T/T_c < 1.05$, all three susceptibilities show rather strong cut-off and volume
effects. Such behavior   is expected due to the first order nature of the phase transition
in pure gauge theory. Outside this region, the fluctuations of longitudinal and the modulus of the Polyakov loop,
show only minimal dependence on $N_\tau$ and $N_\sigma$ in both phases. The transverse susceptibility $\chi_{T}$ in Fig. \ref{fig:sus}, however, is seen to exhibit stronger
$N_\tau$ dependence in the deconfined phase.

\subsection{The ratios of susceptibilities}

The ambiguities from the renormalization scheme can be avoided by considering the ratios of the susceptibilities of the Polyakov loop.
These are particularly interesting observables to identify the deconfinement phase transition \cite{prd}. In the following , we study the influences of dynamical quarks on these ratios.

Figs. \ref{ratioA} and  \ref{ratioT} show the ratios $R_{A}=\chi_{A}/\chi_{L}$ and $R_{T}=\chi_{T}/\chi_{L}$ of different susceptibilities obtained in the SU(3) pure gauge theory. Also shown in this figure are results obtained in lattice QCD with (2+1)-flavor HISQ action and for almost physical quark masses. For $N_f\neq 0$, the temperature  in Figs. \ref{ratioA} and  \ref{ratioT} is normalized to the corresponding pseudo critical temperature at fixed number of flavors 
\cite{ejiri5, hisq}.

\begin{figure*}[!t]
 \includegraphics[width=3.305in]{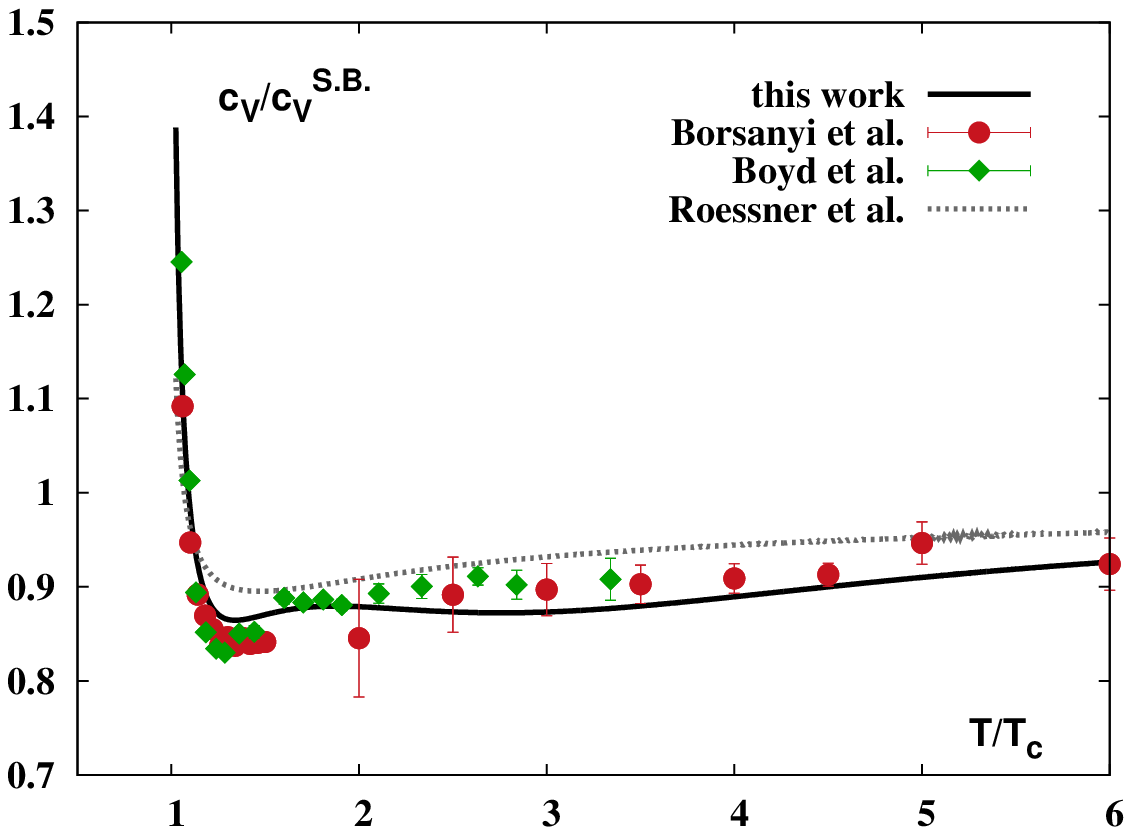}
   \includegraphics[width=3.305in]{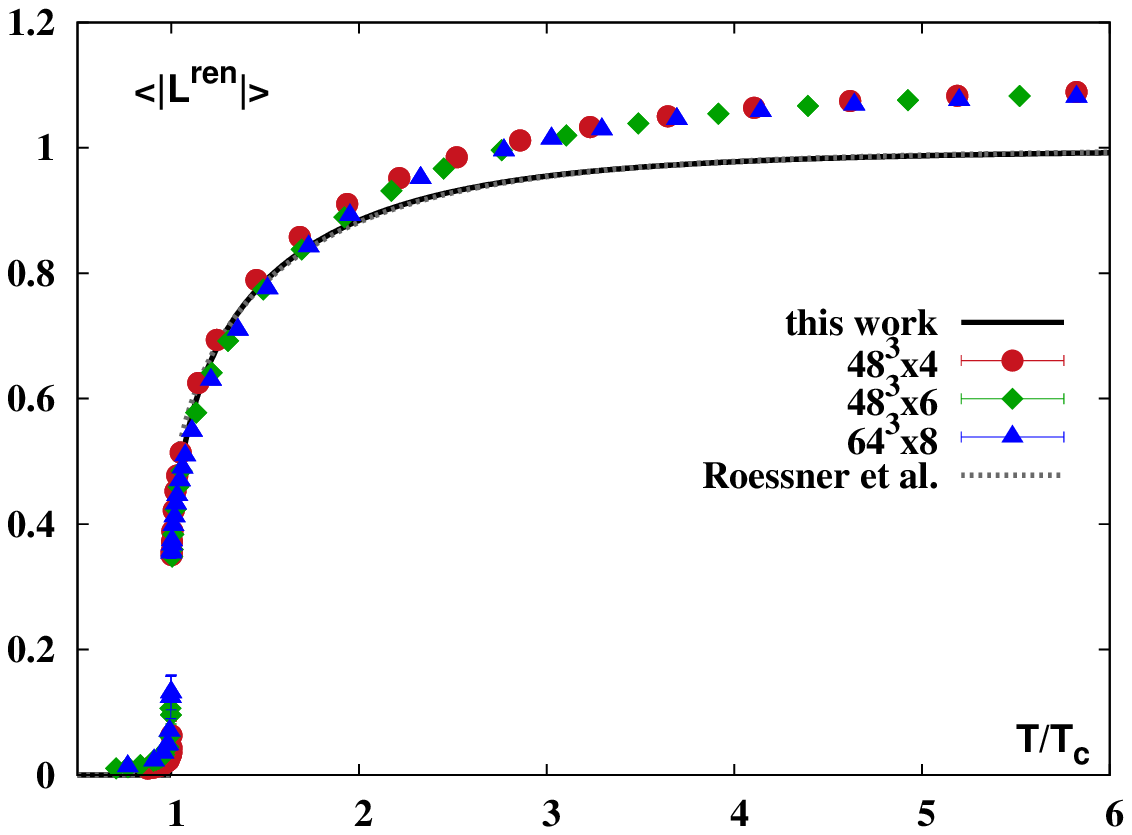}
 \caption{Left-hand figure: The specific heat normalized to its Stefan Boltzmann limit
 obtained from lattice data from Refs. \cite{Boyd:1995zg} and  \cite{Borsanyi:2012ve},  see text.
   Right-hand figure: Points are the thermal averages of the renormalized Polyakov loop as in Fig. \ref{fig:polyakov}. The lines are the Polyakov loop model results discussed in Section III.   }
 \label{fig:SHPL}
\end{figure*}

In the SU(3) pure gauge theory ($N_f=0$),  the ratios $R_A$ and $R_T$  exhibit a $\theta$-like discontinuity at  $T_c$,  and change only weakly with temperature on either side of the transition. Consequently,  the ratios of the Polyakov loop susceptibilities,  can be considered  as an excellent  probe of the phase change.  In the confined phase, their constant values can be deduced from general symmetry arguments,  and the observation that these susceptibilities are dominated by the quadratic terms of the effective action \cite{prd}. In the deconfined phase, the ratio $R_{A}\simeq 1$,   as long as,  the center symmetry is spontaneously broken in the real sector. The properties of $R_T$ above $T_c$ seen in Fig. \ref{ratioT} indicate, that in the SU(3) pure gauge theory the fluctuations  of the longitudinal part of the Polyakov loop exceed that of
  the transverse part,  while keeping the ratio almost $T$-independent.

In the presence of dynamical quarks, the Polyakov loop is no longer an order parameter and stays finite even in the low temperature phase.  Consequently, the ratios of the Polyakov loop susceptibilities are modified due to  explicit breaking of the $\mathcal{Z}_3$ center symmetry. We therefore expect the smoothening of these ratios across the pseudo critical temperature. Indeed Figs. \ref{ratioA} and  \ref{ratioT} show, that in the presence of dynamical quarks, both ratios vary continuously with temperature. We observe that $R_A$ interpolates between the two limiting values set by the pure gauge theory, while the width of crossover region can correlate with the number of flavors and values of their  masses.

The ratio $R_T$ is strongly influenced by the existence of quarks. In addition to the stronger smoothening effect observed, its numerical values in the deconfined phase become temperature dependent,  and deviate substantially from the pure gauge result. We also note  the abrupt change of slopes of $R_T$ and $R_A$ near their respective transition point,  in spite of the fact, that the light quark masses are small.

 A more quantitative investigations of the effects of quarks on the Polyakov loop susceptibilities require further studies on the  system size  and the quark mass dependence of these quantities, as well as their extrapolation
to the continuum and thermodynamic limit.
 We defer such  analysis  to later research.

\section{The Polyakov loop potential}
The non-perturbative aspects of QCD thermodynamics can be studied using effective models, with  quarks and  Polyakov loop as relevant degrees
of freedom \cite{meisinger,Fukushima2004277,PhysRevD.86.014007,Roessner:2006xn,Sasaki:2006ww,PhysRevD.73.014019,Dumitru:2002cf,Dumitru:2003hp,Herbst:2010rf,review}. The parameters of such models are fixed,  so as to reproduce  the lattice results on different observables.
We take advantage of our new lattice data on the Polyakov loop susceptibilities to construct a phenomenological Polyakov loop potential which takes fluctuations into account.

The most commonly used Polyakov loop models are the polynomial \cite{PhysRevD.73.014019} and  the logarithmic potential  \cite{Fukushima2004277,Roessner:2006xn, Sasaki:2006ww},
 the latter includes the contribution of SU(3) Haar measure. While both models are constructed to match the lattice data on  the equation of state and the  average Polyakov loop, they predict  very different results for the Polyakov loop susceptibilities. In particular, the polynomial  model suggests, that  for $ T > T_c$, the ratio $R_T > 1$,   while the logarithmic  model  gives $R_T < 1$. In this regard, the lattice data in Fig. \ref{ratioT} clearly favors the  model  with the Haar measure included. Moreover, as noticed  in Ref.  \cite{Sasaki:2006ww}, the ratio $R_T > 1$,  would lead to  the  negative susceptibility $\chi_{LL} = \chi_{L}-\chi_{T}$, which is unphysical. This is in line with the theoretical expectation,  that the logarithmic  model is preferred,  as it enforces the restriction of the Polyakov loop to the target region \cite{Fukushima2004277}.

Figs. \ref{fig:PM} and \ref{fig:SHPL} show  comparison  of  the logarithmic potential model   from Ref. \cite{Roessner:2006xn}   with the SU(3) lattice data.
The model provides a reasonable description of the  lattice equation of state and the specific heat.
 However, the temperature dependence of the renormalized  Polyakov  loop  is described only  in the vicinity of $T_c$.
This is hardly surprising,  since  the lattice data on  the renormalized Polyakov loop overshoot unity at $T> 3 \, T_c$. For any potential model, that complies to the restriction of the Haar measure, the Polyakov loop can never exceed unity. To tackle this problem in model calculations,  we employ a smooth extrapolation of lattice data, from the temperature range  $ (1.0 - 1.6) \, T_c $,  to unity at high  temperature.  Fig. \ref{fig:SHPL}-right shows such matching of the model results from Ref. \cite{Roessner:2006xn} to the SU(3) lattice data on the renormalized Polyakov loop.

\begin{figure*}[!t]
 \includegraphics[width=3.305in]{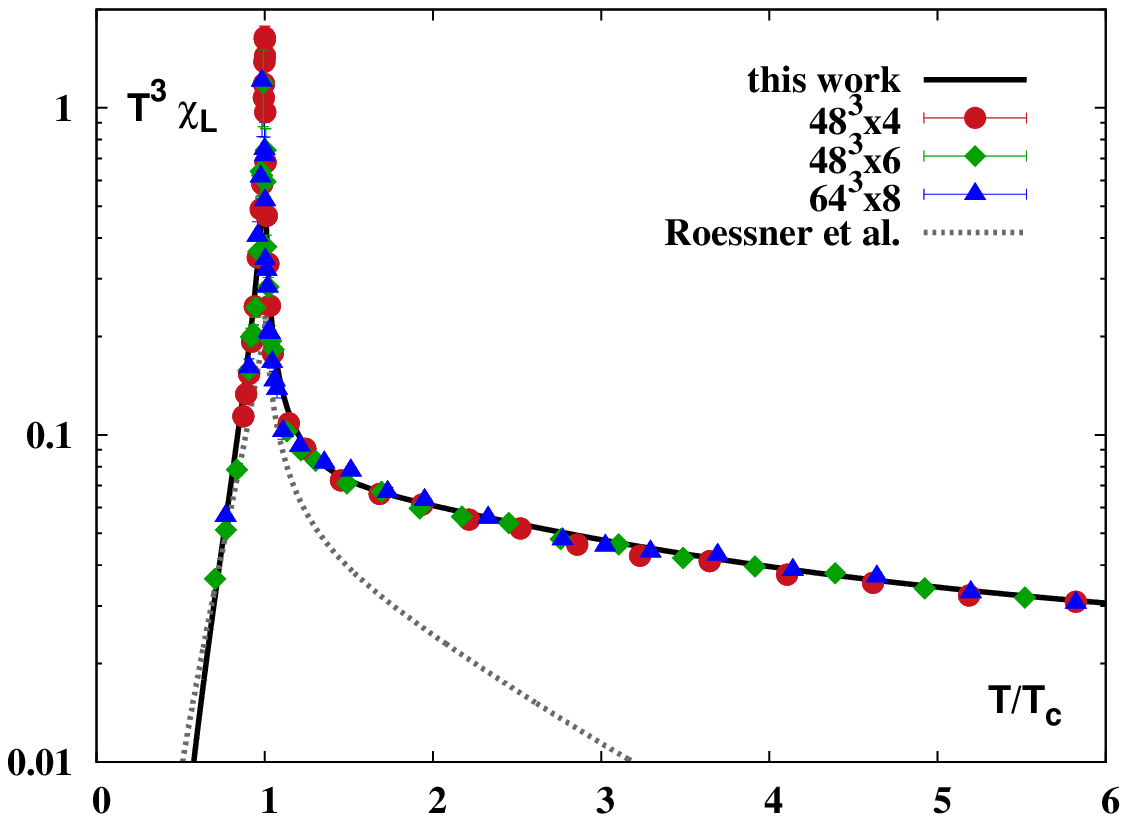}
   \includegraphics[width=3.305in]{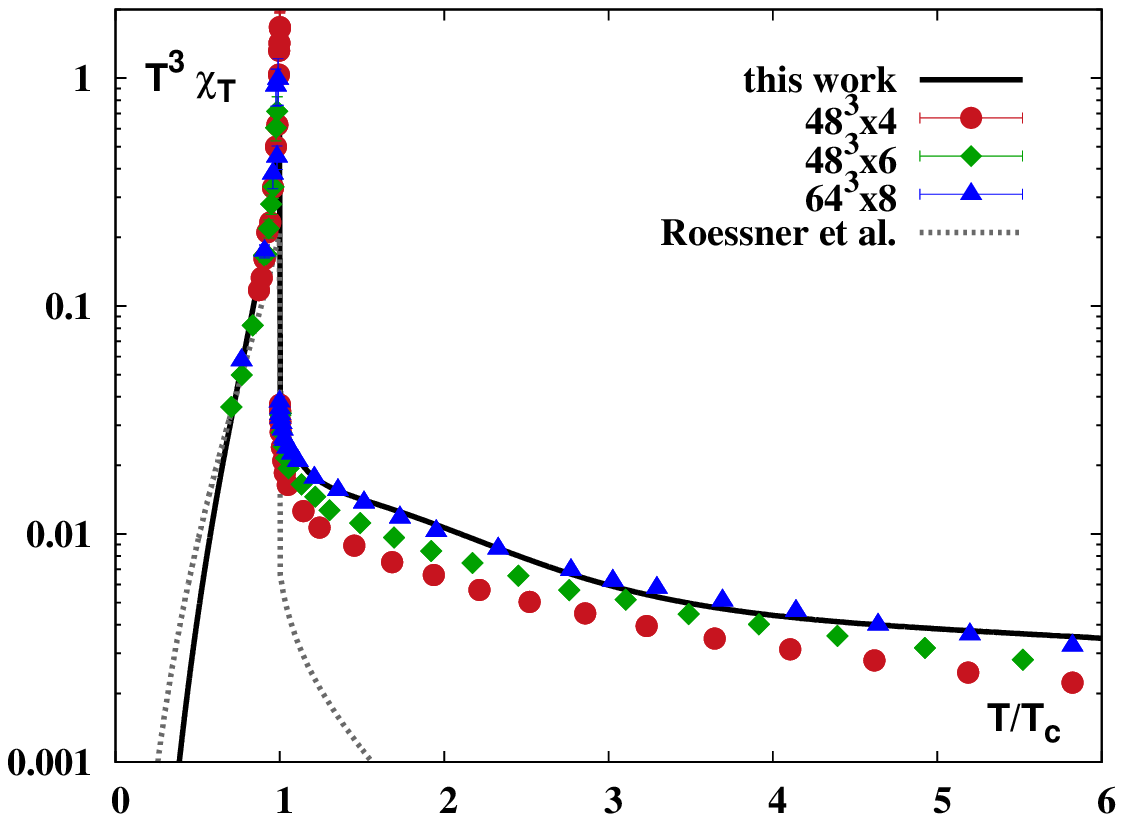}
 \caption{ Lattice data for the longitudinal (left-hand figure)  and the transverse (right-hand figure) Polyakov loop fluctuations from  Fig \ref{fig:sus}.
 The lines are the corresponding potential model results discussed in Section III.}
 \label{fig:m_sus}
\end{figure*}

The  potential from Ref. \cite{Roessner:2006xn} fails to reproduce the temperature dependence of the longitudinal and transverse susceptibility  of the Polyakov loop, as seen in Fig. \ref{fig:m_sus}. Thus,  considerations of the equation of state and $\langle |L|\rangle$ alone,  are not sufficient to describe the pure Yang-Mills thermodynamics. One needs to take into account the effects of fluctuations.

To construct the thermodynamic potential,   which is capable to  quantify  lattice   data on fluctuations,  we consider the    following    $\mathcal{Z}_3$ symmetric  model
\begin{eqnarray}\label{potential}
	\frac{U(L, \bar{L})}{T^4} &= -\frac{1}{2}a(T) \bar{L} L  + b(T) \ln M_{H}(L,\bar L ) \\
&+ \frac{1}{2} c(T) ( L^3 + \bar{L}^3)+ d(T) (\bar{L} L)^2 ,\nonumber
\end{eqnarray}
where  $M_{H}$ is the SU(3) Haar measure,  which is expressed by the Polyakov loop and its conjugate,  as
\begin{eqnarray}
	M_{H} = 1- 6 \bar{L}{L} + 4 (L^3 + \bar{L}^3) -3 (\bar{L}L)^2.
\end{eqnarray}
The restriction of the Polyakov loop to the target region is naturally enforced by the  Haar measure.
For the special case of vanishing $c(T)$ and $d(T)$, the potential in Eq. \eqref{potential} reduces to the same form used in Ref. \cite{Roessner:2006xn}.


The potential  parameters, can be uniquely determined from the lattice data  on the equation of state,   the  Polyakov loop expectation value,  as well as,  its  susceptibilities.
The numerical values  for  model parameters and a description of methods applied are summarized  in the Appendix II.
The  model  predictions on different thermodynamic observables are compared  with lattice data in Figs. \ref{ratioT}-\ref{fig:m_sus}.

 Fig. \ref{fig:PM} shows, that  there is a very satisfactory description of lattice results on the thermodynamic pressure and the interaction measure,  up to  very high  temperatures.
Our model parameters are tuned to describe the most recent \cite{Borsanyi:2012ve},  rather than previous  \cite{Boyd:1995zg}, lattice data on the thermodynamic pressure.   This yields small differences between our and the  previous \cite{Roessner:2006xn} Polyakov loop model results.

The comparison of model predictions  with lattice
data for heat capacity $c_V$, normalized to its Stefan Boltzmann limit $ c_V^{SB}=({32 \pi^2}/{15})T^3 $,  is shown in Fig. \ref{fig:SHPL}.
The lattice
data for $c_V(T)$ were extracted from the respective continuum pressure
results. The derivatives were calculated  by taking central differences between
data points. The errors were estimated by $\mathcal{O}[h^2]$, where $h =
\delta T/T$ and $\delta T$ is the difference between the grids around the
specific point. The $c_V$ exhibits a shallow minimum above $T_c$,  which  is  reproduced by the
present Polyakov loop model quite satisfactory.

The essential difference between our model in Eq. \eqref{potential} and the previously proposed logarithmic potential,  lies in the prediction for the Polyakov loop fluctuations. Figs. \ref{fig:m_sus} shows, that both the  longitudinal and the transverse susceptibility are very well described by the new Polyakov loop potential, while the logarithmic model tends to underestimate both susceptibilities.


\section{Conclusions}

We have calculated the Polyakov loop susceptibilities in the SU(3) lattice gauge theory for different number of quark flavors, $N_f=0$  and
$N_f=(2+1)$. For (2+1)-flavor QCD, the susceptibilities were calculated from the Polyakov loop results from \cite{hisq}, where the HISQ action with almost physical strange quark mass and $m_{u,d} = m_s/20$ was used. For all cases, the extrapolations to the continuum and thermodynamic limit have  not been  performed yet.


We have discussed   the temperature dependence of the
longitudinal $\chi_L$, the  transverse $\chi_T$ and the absolute value $\chi_A$ of the Polyakov loop fluctuations.  We have analyzed   general
  properties
  of the  Polyakov loop susceptibilities
  in relation to the color group structure and motivated  their  ratios, $R_{A}=\chi_{A}/\chi_{L}$ and $R_{T}=\chi_{T}/\chi_{L}$,   as relevant observables to probe deconfinement.

A  remarkable feature  of different  ratios of the Polyakov loop susceptibilities   is their strong sensitivity to a phase change
in a system.  In the SU(3) pure gauge theory the ratios of susceptibilities  show  discontinuity  at the critical point   and exhibit a very weak, but different,  temperature dependence in the confined and deconfined phase. The  explicit breaking of the $\mathcal{Z}_3$ center symmetry in QCD, due to quark fields,  modifies this property.

The $R_A$,  and in particular $R_T$ ratios are substantially smoothened, yet still display interesting features related to the deconfinement.
The $R_A$  converges  to the asymptotic values found in a pure gauge theory, both   in the confined and the deconfined phase.
Whereas, the  $R_T$   converges only to  the  low temperature limit of   a  pure gauge theory. At high temperatures and  for light quarks,   it differs  substantially from the results obtained in   a pure gauge theory.


We have used our  data on the Polyakov loop  susceptibilities
to construct an effective Polyakov loop potential, which is consistent  with all lattice results over a broad range of temperatures.  We have shown, that the incorporation of fluctuations  is important to  describe thermodynamics of a pure gauge theory within the effective approach.
The Polyakov loop potential  developed in this work is open to a more realistic
effective description of QCD thermodynamics with quarks.

\acknowledgments
We acknowledge stimulating discussions with J\"urgen Engels, Frithjof  Karsch and Helmut Satz.
We are grateful to Peter Petreczky, Alexei Bazavov and the HotQCD Collaboration for providing us with data for the Polyakov loop.
P.M.L. acknowledges the support of the Frankfurt Institute for Advanced Studies (FIAS).
B.F. is supported in part by the Extreme Matter Institute EMMI.
K.R. acknowledges partial support of the Polish Ministry of National
Education (NCN). The work of C.S. has been partly supported
by the Hessian LOEWE initiative through the Helmholtz
International Center for FAIR (HIC for FAIR). The numerical calculations
have been performed on the Bielefeld GPU Cluster.

\section*{Appendix I}\label{app1}

In Table \ref{tab:table1}, we summarize our lattice results on the Polyakov loop and its susceptibilities
for SU(3) pure gauge theory on $64^3 \times 8$ lattice.

\begin{table}
\begin{center}
\begin{tabular}{|c|c|c|c|c|}
      \hline
       $T/T_c$ & $ <|L^{\rm ren}|> $ & $T^3 \chi_A$ & $T^3 \chi_L$ & $T^3 \chi_T$ \\
       \hline
       $0.768$ &  $0.0132(1)  $ & $0.0245(4)$ & $0.0564(12)$ & $ 0.0576(16)  $ \\
       \hline
       $0.908$ &  $0.0225(6)  $ & $0.074(4)$ & $0.161(10) $ & $ 0.174(11)   $ \\
       \hline
       $0.958$ &  $0.0349(17) $ & $0.178(24)$ & $0.405(44) $ & $ 0.379(53)   $ \\
       \hline
       $0.977$ &  $0.0478(36) $ & $0.344(42)$ & $0.615(86) $ & $ 0.92(17)    $ \\
       \hline
       $0.983$ &  $0.0484(62) $ & $0.56(25)$ & $1.21(39)  $ & $ 0.450(53)   $ \\
       \hline
       $0.990$ &  $0.0699(62) $ & $0.79(16)$ & $2.14(47)  $ & $ 0.99(23)    $ \\
       \hline
       $0.997$ &  $0.124(34)  $ & $6.(2)$ & $4.40(85)  $ & $ 10.(3)      $ \\
       \hline
       $0.998$ &  $0.132(27)  $ & $7.(2)$ & $12.(4)    $ & $ 4.(1)       $ \\
       \hline
       $1.000$ &  $0.354(8)   $ & $0.74(16)$ & $0.75(16)  $ & $ 0.038(2)    $ \\
       \hline
       $1.001$ &  $0.357(5)   $ & $0.71(10)$ & $0.71(10)  $ & $ 0.035(1)    $ \\
       \hline
       $1.002$ &  $0.367(3)   $ & $0.520(65)$ & $0.521(65) $ & $ 0.033(1)    $ \\
       \hline
       $1.004$ &  $0.374(3)   $ & $0.343(20)$ & $0.343(20) $ & $ 0.032(1)    $ \\
       \hline
       $1.010$ &  $0.398(3)   $ & $0.319(28)$ & $0.319(28) $ & $ 0.0305(6)   $ \\
       \hline
       $1.017$ &  $0.412(2)   $ & $0.283(30)$ & $0.283(30) $ & $ 0.0286(6)   $ \\
       \hline
       $1.024$ &  $0.432(2)   $ & $0.207(13)$ & $0.207(13) $ & $ 0.0263(4)   $ \\
       \hline
       $1.031$ &  $0.446(1)   $ & $0.204(12)$ & $0.204(12) $ & $ 0.0257(3)   $ \\
       \hline
       $1.045$ &  $0.469(1)   $ & $0.166(7)$ & $0.166(7)  $ & $ 0.0239(2)   $ \\
       \hline
       $1.059$ &  $0.490(1)   $ & $0.147(10)$ & $0.147(10) $ & $ 0.0239(4)   $ \\
       \hline
       $1.073$ &  $0.509(1)   $& $0.138(8)$  & $0.138(8)  $ & $ 0.0224(3)   $\\
       \hline
       $1.109$ &  $0.548(1)   $ & $0.102(5)$ & $0.102(5)  $ & $ 0.0209(3)   $ \\
       \hline
       $1.211$ &  $0.630(1)   $ & $0.092(3)$ & $0.092(3)  $ & $ 0.0176(2)   $ \\
       \hline
       $1.354$ &  $0.709(1)   $ & $0.082(2)$ & $0.082(2)  $ & $ 0.0155(2)   $ \\
       \hline
       $1.512$ &  $0.775(0)   $ & $0.078(2)$ & $0.078(2)  $ & $ 0.0137(2)   $ \\
       \hline
       $1.731$ &  $0.842(0)   $ & $0.067(2)$ & $0.067(2)  $ & $ 0.0118(2)   $ \\
       \hline
       $1.951$ &  $0.892(0)   $ & $0.063(2)$ & $0.063(2)  $ & $ 0.0103(1)   $ \\
       \hline
       $2.328$ &  $0.951(0)   $ & $0.056(1)$ & $0.056(1)  $ & $ 0.0086(1)   $ \\
       \hline
       $2.772$ &  $0.995(0)   $ & $0.048(1)$ & $0.048(1)  $ & $ 0.00692(1)  $ \\
       \hline
       $3.026$ &  $1.014(0)   $ & $0.046(1)$ & $0.046(1)  $ & $ 0.00622(1)  $ \\
       \hline
       $3.294$ &  $1.029(0)   $ & $0.044(1)$ & $0.044(1)  $ & $ 0.00578(1)  $ \\
       \hline
       $3.694$ &  $1.045(0)   $ & $0.043(1)$ & $0.043(1)  $ & $ 0.00509(1)  $ \\
       \hline
       $4.140$ &  $1.058(0)   $ & $0.039(1)$ & $0.039(1)  $ & $ 0.00458(0)  $ \\
       \hline
       $4.639$ &  $1.068(0)   $ & $0.037(1)$ & $0.037(1)  $ & $ 0.00399(1)  $ \\
       \hline
       $5.198$ &  $1.076(0)   $ & $0.033(1)$ & $0.033(1)  $ & $ 0.00361(0)  $ \\
       \hline
       $5.823$ &  $1.081(0)   $ & $0.031(1)$ & $0.031(1)  $ & $ 0.00322(0)  $ \\
       \hline
\end{tabular}
\end{center}
\caption{The lattice results for the Polyakov loop and its susceptibilities, defined in Eqs. \eqref{eq:chiA}--\eqref{eq:chiT},   obtained in the SU(3) pure gauge theory on $64^3 \times 8$ lattice.}
\label{tab:table1}
\end{table}

\section*{Appendix II}\label{app2}

To fix  parameters of the phenomenological gluon  potential introduced in Eq. \eqref{potential}, we first express  $U(L,\bar L)$  in terms of longitudinal $L_L$ and transverse $L_T$ parts of the Polyakov loop,
\begin{eqnarray}\label{potri}
	\frac{U(L_L,L_T)}{T^4} &=& -\frac{a(T)}{2} (L_{L}^2 + L_{T}^2)  + b(T) \ln M_{H}\\
					     &+& c(T) (L_{L}^3 - 3 L_{L} L_{T}^2) + d(T) ( L_{L}^2 + L_{T}^2 )^2 \nonumber
\end{eqnarray}
with the Haar measure,
\begin{eqnarray}
M_{H} =& 1- 6 (L_{L}^2 + L_{T}^2) + 8 ( L_{L}^3  - 3 L_{L} L_{T}^2) \\
&-3 (L_{L}^2 + L_{T}^2)^2,\nonumber
\end{eqnarray}
and the parameters $a(T),b(T),c(T)$ and $d(T)$.

The expectation values of both components of the Polyakov loop are obtained by solving the gap equations,
\begin{eqnarray}
\frac{\partial {U}(L_{L}, L_{T})}{\partial L_{L}} = 0, ~~\frac{\partial {U}(L_{L}, L_{T})}{\partial L_{T}} = 0.	
\end{eqnarray}

Focusing on the real sector, we denote the solutions of the gap equations as, $L_{T} = 0$
and $L_{L} = L_0$. The pressure is then given by

\begin{eqnarray}
        \frac{P}{T^4} = -{\frac{U}{T^4}}(L_{L} \rightarrow L_0, L_{T} \rightarrow 0; \, T).
\end{eqnarray}

To calculate the susceptibilities, we construct the curvature tensor,
\begin{eqnarray}
	\hat{\mathcal{C}}_{ik} = \frac{1}{T^4}\frac{\delta^2 {U}}{\delta L_i \delta L_k},
\end{eqnarray}
which is a $2 \times2$ matrix,  with $L_k = \{L_{L}, L_{T} \}$.
The susceptibility matrix is defined as,
$
T^3 \hat{\chi} = \hat{\mathcal{C}}^{-1}.
$

In the real sector, the susceptibility matrix is diagonal. The explicit expressions for the longitudinal and transverse Polyakov loop susceptibilities read

\begin{eqnarray}
\left(T^3 \chi_{L}\right)^{-1} &= -a(T) + b(T)(-\frac{3}{(L_0-1)^2} \\
&- \frac{9}{(1+3 L_0)^2}) + 6 L_0 c(T) + 12 d(T) L_0^2 , \nonumber
\end{eqnarray}

\begin{eqnarray}
\left( T^3 \chi_{T} \right)^{-1} &= -a(T) + b(T)\frac{12 (1 + L_0 (4 + L_0))}{(L_0-1)^3 (1 + 3 L_0)} \\
&- 6 L_0 c(T) + 4 d(T) L_0^2. \nonumber
\end{eqnarray}

These two expressions, together with the gap and pressure equations, are sufficient to determine the
potential parameters at different temperatures, having the lattice data for the thermodynamics observables.

The unique matching of model parameters with lattice data is only possible in the symmetry broken phase. In the confined phase, the Polyakov loop vanishes and the set of equations is no longer linear independent. The lattice susceptibilities data constrain only the quadratic terms in the effective potential and there is no restriction on parameters $c(T)$ and $d(T)$. In principle, they can be fixed by lattice data on higher order cumulants of the Polyakov loops, which are neglected in the present study.


The temperature dependence of model parameters in Eq. \eqref{potri} can be expressed by the following parametrizations,

\begin{eqnarray}
	\label{par1}
	x(T) = 
 (x_1 + x_2/t + x_3/t^2) (1 + x_4/t + x_5/t^2)^{-1}
\end{eqnarray}
for $x=(a,c,d)$, whereas $b(T)$ is parameterized as

\begin{eqnarray}
	\label{par2}
	b(T) = {b_1}{t^{-b_4}} (1-e^{b_2/t^{b_3}}),
\end{eqnarray}
where $t = {T}/{T_c}$ and $T_c$ is the deconfinement temperature in the SU(3) lattice gauge theory.

Table \ref{tab:table2} summarizes our numerical results for the model parameters. The model potential is constructed to quantitatively describe SU(3) lattice data of different thermodynamic observables, including results for the Polyakov loop susceptibilities.

\begin{table}
\
\begin{center}
\begin{tabular}{|c|c|c|c|c|c|c|c|c|c|c|}
    \hline
      $ a_1 $ & $a_2$ & $a_3$ & $a_4$ & $a_5$\\
    \hline
      $ -44.14 $ & $ 151.4 $ & $ -90.0677 $ & $ 2.77173 $ & $ 3.56403 $ \\
    \hline
     $b_1$ & $b_2$  & $b_3$ & $b_4$ & \\
    \hline
     $ -0.32665 $ & $ -82.9823 $ & $ 3.0 $ & $ 5.85559 $ &\\
    \hline
     $c_1$ & $c_2$ & $c_3$ & $c_4$ & $c_5$\\
    \hline
     $-50.7961 $ & $114.038 $ & $-89.4596 $ & $ 3.08718 $ & $6.72812  $  \\
    \hline
     $d_1$ & $d_2$ & $d_3$ & $d_4$ & $d_5$\\
    \hline
    $27.0885 $ & $-56.0859 $ & $71.2225 $ & $2.9715 $ & $ 6.61433 $ \\
    \hline
\end{tabular}
\end{center}
\caption{The numerical values of the parameters $a(T), b(T), c(T)$ and $d(T)$  for the Polyakov loop potential in Eq. \eqref{potential},  under the parametrizations introduced in Eqs. \eqref{par1} and \eqref{par2}. }
\label{tab:table2}
\
\end{table}

We have employed lattice results for thermodynamic pressure and interaction measure from Ref. \cite{Borsanyi:2012ve}, while the results for the Polyakov loop  were taken  from  Ref. \cite{Kaczmarek:2002mc}.  The data for the longitudinal and transverse Polyakov loop susceptibilities, incorporated into our effective potential, were introduced in Section II.

Currently, the lattice data on the Polyakov loop susceptibilities are available only in the temperature interval $ 0.8 < T/T_c < 6$. Our model parameters, presented in Table \ref{tab:table2}, are inherently restricted to this range.


\end{document}